\documentclass[sort&compress,final]{aipproc}\layoutstyle{6x9}
\begin{document}\title{Meson--Photon Transition Form Factors}
\author{Irina Balakireva}{address={SINP, Moscow State University,
119991 Moscow, Russia}}\author{Wolfgang Lucha}{address={Institute
for High Energy Physics, Austrian Academy of
Sciences,\\Nikolsdorfergasse 18, A-1050 Vienna,
Austria}}\author{Dmitri Melikhov}{address={Institute for High
Energy Physics, Austrian Academy of Sciences,\\Nikolsdorfergasse
18, A-1050 Vienna, Austria},altaddress={Faculty of Physics,
University of Vienna, Boltzmanngasse 5, A-1090 Vienna, Austria}}
\begin{abstract}We present the results of our recent analysis of
the meson--photon transition form factors $F_{P\gamma}(Q^2)$ for
the pseudoscalar mesons $P = \pi^0,\eta,\eta',\eta_c$, using the
local-duality version of QCD sum rules.\end{abstract}
\maketitle

\section{Introduction}The processes $\gamma^*\,\gamma^*\to P$ with
$P=\pi^0,\eta,\eta',\eta_c$ are of great interest for our
understanding of QCD and of the meson structure. In recent years,
extensive experimental information on these processes has become
available \cite{cello-cleo,babar1,babar,babar2010,belle}.

The corresponding amplitude contains only one form factor,
$F_{P\gamma\gamma}(q_1^2,q_2^2)$:
\begin{eqnarray}
\langle\gamma^*(q_1)\gamma^*(q_2)|P(p)\rangle={\rm i}
\epsilon_{\varepsilon_1\varepsilon_2
q_1q_2}F_{P\gamma\gamma}(q_1^2,q_2^2).
\end{eqnarray}
A QCD factorization theorem predicts this form factor at
asymptotically large spacelike momentum transfers
$q_1^2\equiv-Q_1^2\le0$, $q_2^2\equiv-Q_2^2\le0$ \cite{brodsky}:
\begin{eqnarray}
F_{P\gamma\gamma}(Q_1^2,Q_2^2)\to2e_c^2\int\limits_0^1\frac{{\rm
d}\xi\phi^{\rm ass}_P(\xi)}{Q_1^2\xi+Q_2^2(1-\xi)},\qquad
\phi^{\rm ass}_P(\xi)=6f_P\xi(1-\xi).
\end{eqnarray}
Hereafter, we use the notation $Q^2\equiv Q_2^2$ and
$0\le\beta\equiv Q_1^2/Q_2^2 \le 1$ (that is, $Q_2^2$ is
the~larger virtuality). For the experimentally relevant kinematics
$Q_1^2\approx0$ and $Q_2^2\equiv Q^2$, for instance, the
pion--photon transition form factor takes the form
\begin{eqnarray}
Q^2F_{\pi\gamma}(Q^2)\to\sqrt{2}f_\pi,\qquad
f_\pi=0.130\;\mbox{GeV}.
\end{eqnarray}
Similar relations arise for $\eta$ and $\eta'$ after taking into
account the effects of meson~mixing.

\section{Dispersive sum rules for the $\gamma^*\,\gamma^*\to P$ form factor}
The starting point for a QCD sum-rule analysis of the
$\gamma^*\,\gamma^*\to P$ transition form factor~is the amplitude
\begin{eqnarray}
\label{AVV}\langle 0| j_{\mu}^5|\gamma^*(q_2)\gamma^*(q_1)\rangle
=e^2T_{\mu\alpha\beta}(p|q_1,q_2)\,\varepsilon^\alpha_1\varepsilon^\beta_2,
\qquad p=q_1+q_2,
\end{eqnarray}
where $\varepsilon_{1,2}$ are the relevant photon polarization
vectors. This amplitude is considered for $-q_1^2\equiv Q_1^2\ge0$
and $-q_2^2\equiv Q_2^2\ge0$. Its general decomposition contains
four independent Lorentz structures (see e.g.\
Refs.~\cite{blm2011,lm2011}) but for our purpose only one
structure is needed:
\begin{eqnarray}
\label{F} T_{\mu\alpha\beta}(p|q_1,q_2)= p_\mu
\epsilon_{\alpha\beta q_1 q_2}{\rm i}F(p^2,Q_1^2,Q_2^2)+\cdots.
\end{eqnarray}
The corresponding invariant amplitude $F(p^2,Q_1^2, Q_2^2)$
satisfies the spectral representation in $p^2$ at fixed $Q_1^2$
and $Q_2^2$
\begin{eqnarray}
F(p^2,Q_1^2,Q_2^2)=\frac{1}{\pi}\int\limits_{s_{\rm
th}}^\infty\frac{{\rm d}s}{s-p^2}\,\Delta(s,Q_1^2,Q_2^2),
\end{eqnarray}
where $\Delta(s,Q_1^2,Q_2^2)$ is the physical spectral density and
$s_{\rm th}$ denotes the physical threshold.

Perturbation theory yields the spectral density as a series
expansion in powers of $\alpha_s$:
\begin{eqnarray}
\Delta_{\rm pQCD}(s,Q_1^2,Q_2^2|m)=\Delta^{(0)}_{\rm
pQCD}(s,Q_1^2,Q_2^2|m)+\frac{\alpha_s}{\pi}\Delta^{(1)}_{\rm
pQCD}(s,Q_1^2,Q_2^2|m)+\cdots,
\end{eqnarray}
where $m$ is the mass of the quark propagating in the loop. The
lowest-order contribution, $\Delta^{(0)}_{\rm
pQCD}(s,Q_1^2,Q_2^2|m)$, corresponding to a one-loop triangle
diagram with one axial current and two vector currents at the
vertices, is well-known \cite{1loop}. The two-loop $O(\alpha_s)$
correction to the spectral density was found to vanish
\cite{2loop}. Higher-order corrections are unknown.

The physical spectral density differs dramatically from
$\Delta_{\rm pQCD}(s,Q_1^2,Q_2^2)$ in the low-$s$ region; it
contains the meson pole and the hadronic continuum. For instance,
in the $I=1$ channel, one has
\begin{eqnarray}
\label{Aps2} \Delta(s,Q_1^2,Q_2^2) = \pi \delta(s - m_{\pi}^2)\,
\sqrt{2} f_{\pi}\,F_{\pi\gamma\gamma}(Q^2_1,Q_2^2) + \theta(s -
s_{\rm th})\,\Delta^{I=1}_{\rm cont}(s,Q^2_1,Q_2^2).
\end{eqnarray}
The method of QCD sum rules allows one to relate the properties of
the ground states to the spectral densities of QCD correlators.
The following steps are conventional within the QCD sum-rule
method \cite{lms1,lms2}: equate the QCD and the physical
representations for $F(p^2,Q_1^2,Q_2^2)$; then perform the Borel
transform $p^2\to \tau,$ which suppresses the hadronic continuum;
in order to kill then potentially dangerous nonperturbative power
corrections which may rise with $Q^2$, take the local-duality (LD)
limit $\tau=0$ \cite{ld}; finally, implement quark--hadron duality
in a standard way as low-energy cut on the spectral
representation, in order to arrive at the following expression for
the ground-state transition form factor:
\begin{eqnarray}
\label{ldsr} \pi f_P
F_{P\gamma\gamma}(Q_1^2,Q_2^2)=\int\limits_{4m^2}^{s_{\rm
eff}(Q_1^2,Q_2^2)}{\rm d}s \,\Delta_{\rm pQCD}(s,Q_1^2,Q_2^2|m).
\end{eqnarray}
All details of the nonperturbative-QCD dynamics are contained in
the effective threshold $s_{\rm eff}(Q_1^2,Q_2^2)$. The
formulation of reliable criteria for fixing effective thresholds
proves to be highly nontrivial \cite{lms1}.

At large $Q_2^2\equiv Q^2\to\infty$ and fixed ratio $\beta\equiv
Q_1^2/Q_2^2$, the effective threshold $s_{\rm eff}(Q_1^2,Q_2^2)$
may be determined by suitable matching to the asymptotic pQCD
factorization formula. From this, one finds that, in the general
case $m\ne 0$, $s_{\rm eff}(Q^2\to\infty,\beta)$ depends on
$\beta$. The only exception to this is the case of massless
fermions, $m=0$: in this case the asymptotic factorization formula
is reproduced for any $\beta$ if one sets $s_{\rm
eff}(Q^2\to\infty,\beta)=4\pi^2 f_\pi^2.$ The LD \emph{model\/}
for the transition form factor emerges when one \emph{assumes\/}
that, at finite values of $Q^2$, $s_{\rm eff}(Q^2,\beta)$ may be
sufficiently well approximated by its value for $Q^2\to\infty$,
that~is,
\begin{eqnarray}
s_{\rm eff}(Q^2,\beta)=s_{\rm eff}(Q^2\to\infty,\beta).
\end{eqnarray}
Introducing the abbreviation $F_{P\gamma}(Q^2)\equiv
F_{P\gamma\gamma}(0,Q^2)$ for the pseudoscalar-meson--photon
transition form factor, its LD expression for $Q_1^2=0$ and $m=0$
reads, in the single-flavour case,
\begin{eqnarray}
\label{srfp} F_{P\gamma}(Q^2)=\frac{1}{2\pi^2f_P}\frac{s_{\rm
eff}(Q^2)}{s_{\rm eff}(Q^2)+Q^2}.
\end{eqnarray}
Independently of the behaviour of $s_{\rm eff}(Q^2)$ at $Q^2\to
0$, $F_{P\gamma}(Q^2=0)$ is related to the axial anomaly
\cite{blm2011}.

\section{The transition $\gamma\,\gamma^*\to P$ in quantum mechanics}
The accuracy of the LD model for the effective threshold may be
estimated in quantum mechanics. There, the form factor may be
found exactly by some numerical solution \cite{Lucha98} of the
Schr\"odinger equation. From this, the \emph{exact\/} effective
threshold may be calculated: for any given experimental or
theoretical form factor, the corresponding exact effective
threshold is defined as the quantity that reproduces this form
factor by a LD sum rule~(\ref{ldsr}).

\begin{figure}[hb]\begin{tabular}{cc}
\includegraphics[width=7.23cm]{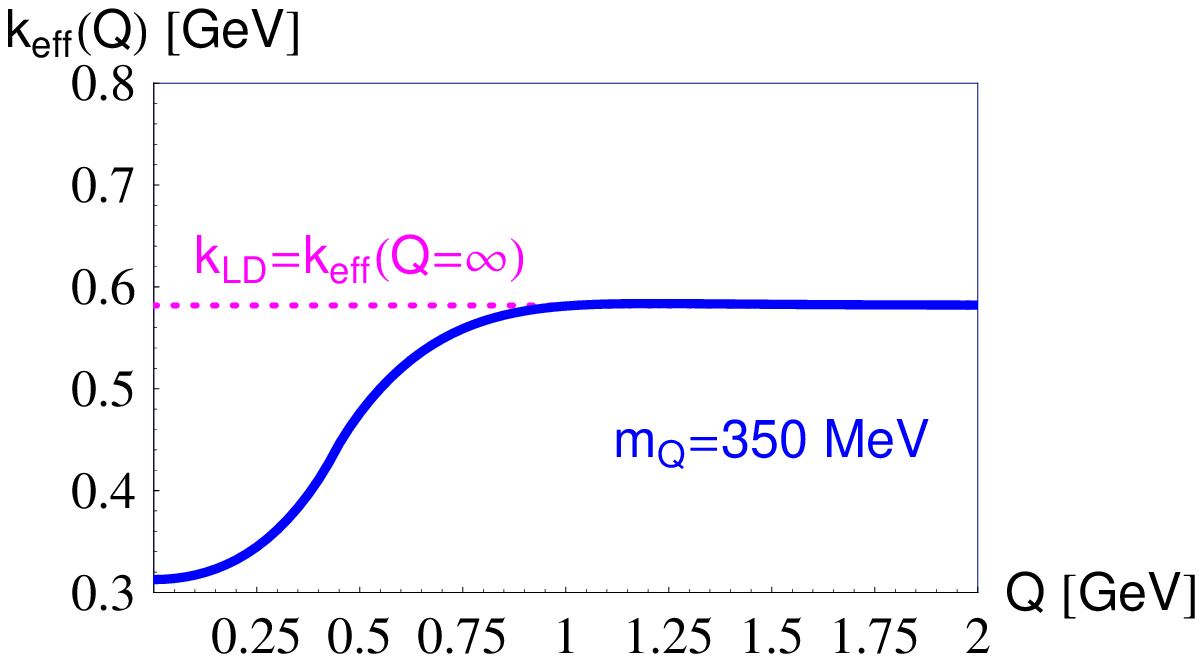}&
\includegraphics[width=7.23cm]{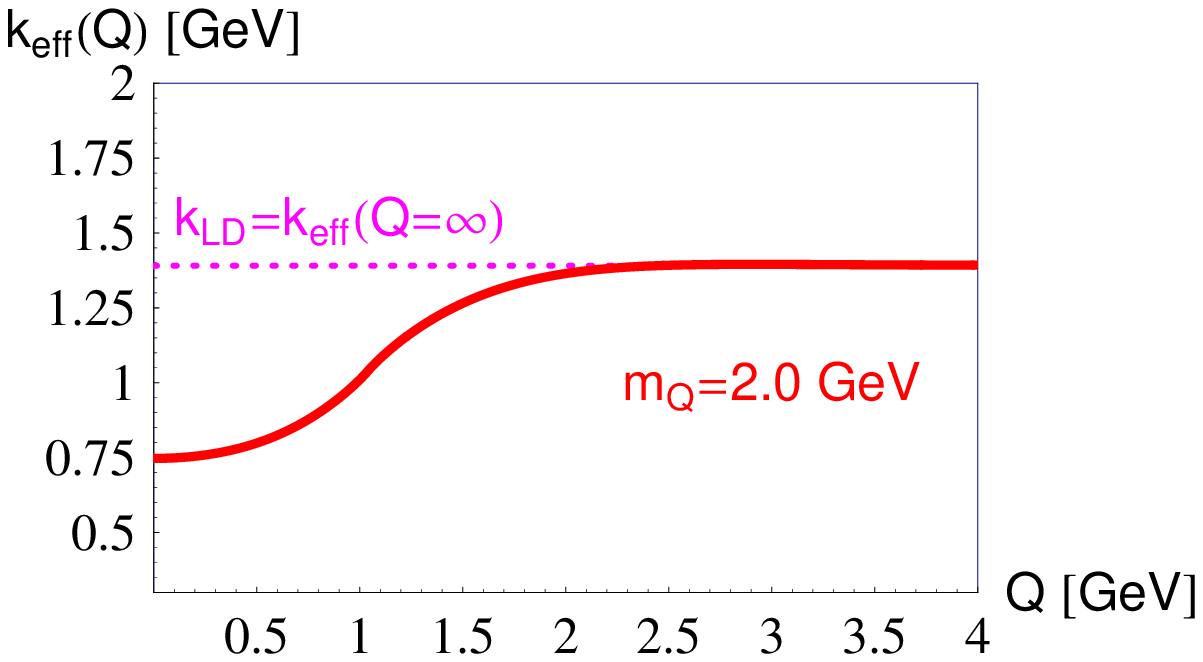}
\end{tabular}\caption{The exact effective threshold in quantum
mechanics, $k_{\rm eff}(Q)$, for two different values of the
nonrelativistic constituent quark mass $m_Q$.}\label{Fig:1}
\end{figure}

The result from a quantum-mechanical model with a
harmonic-oscillator potential \cite{blm2011} is shown in
Fig.~\ref{Fig:1}. For ``light'' quarks, the LD threshold gives a
very good approximation to the exact threshold for
$Q>1$--$1.5\;\mbox{GeV}$. For ``charm'' quarks, the local-duality
model works for $Q>2$--$3\;\mbox{GeV}$. The accuracy of the LD
approximation further increases with~$Q$ in this region.

\section{$\gamma^*\,\gamma^*\to\eta_c$ form factor}
\begin{figure}[hb]\begin{tabular}{cc}
\includegraphics[width=7.2cm]{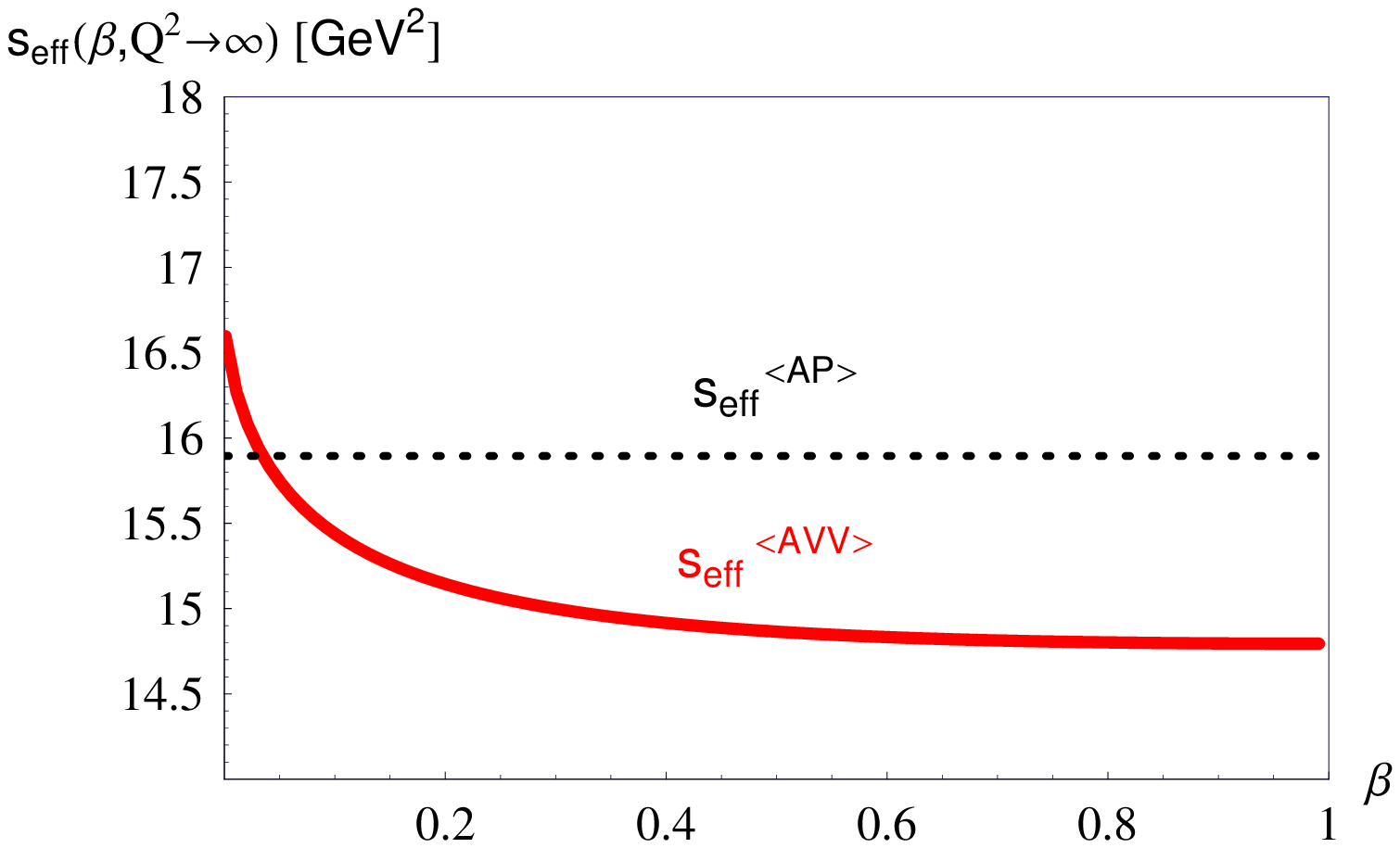}&
\includegraphics[width=7.2cm]{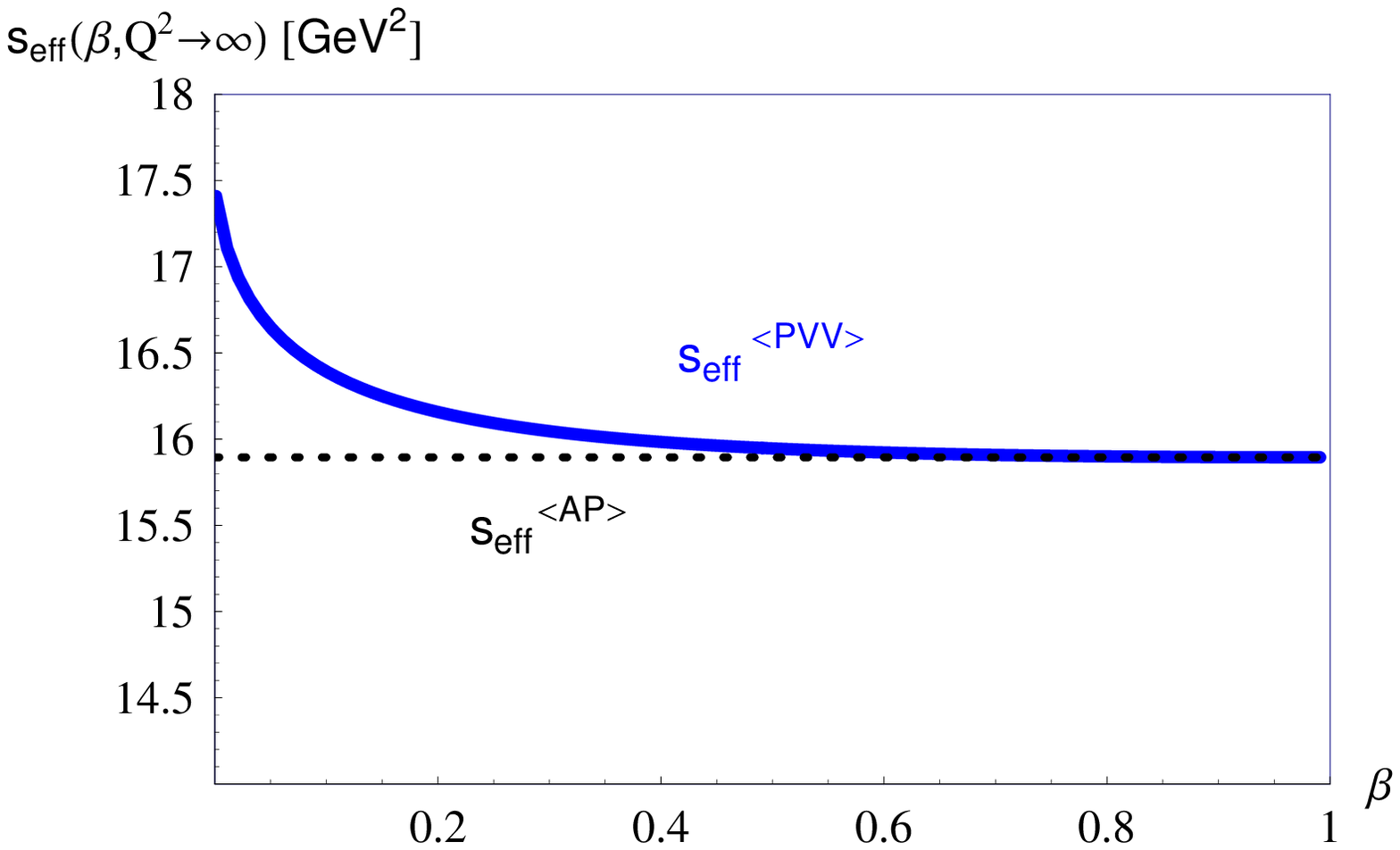}\\(a)&(b)\\[1ex]
\includegraphics[width=7.2cm]{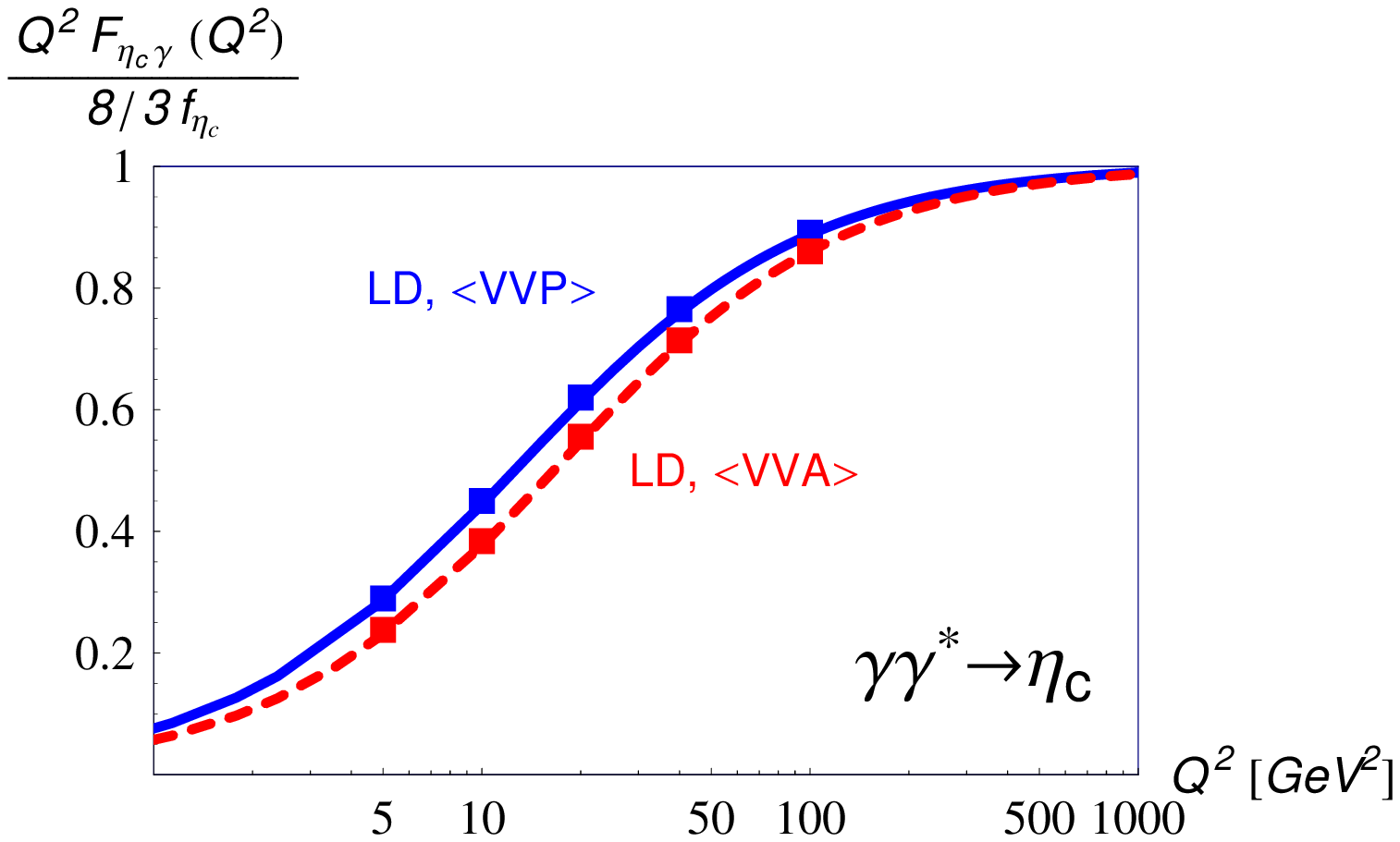}&
\includegraphics[width=7.2cm]{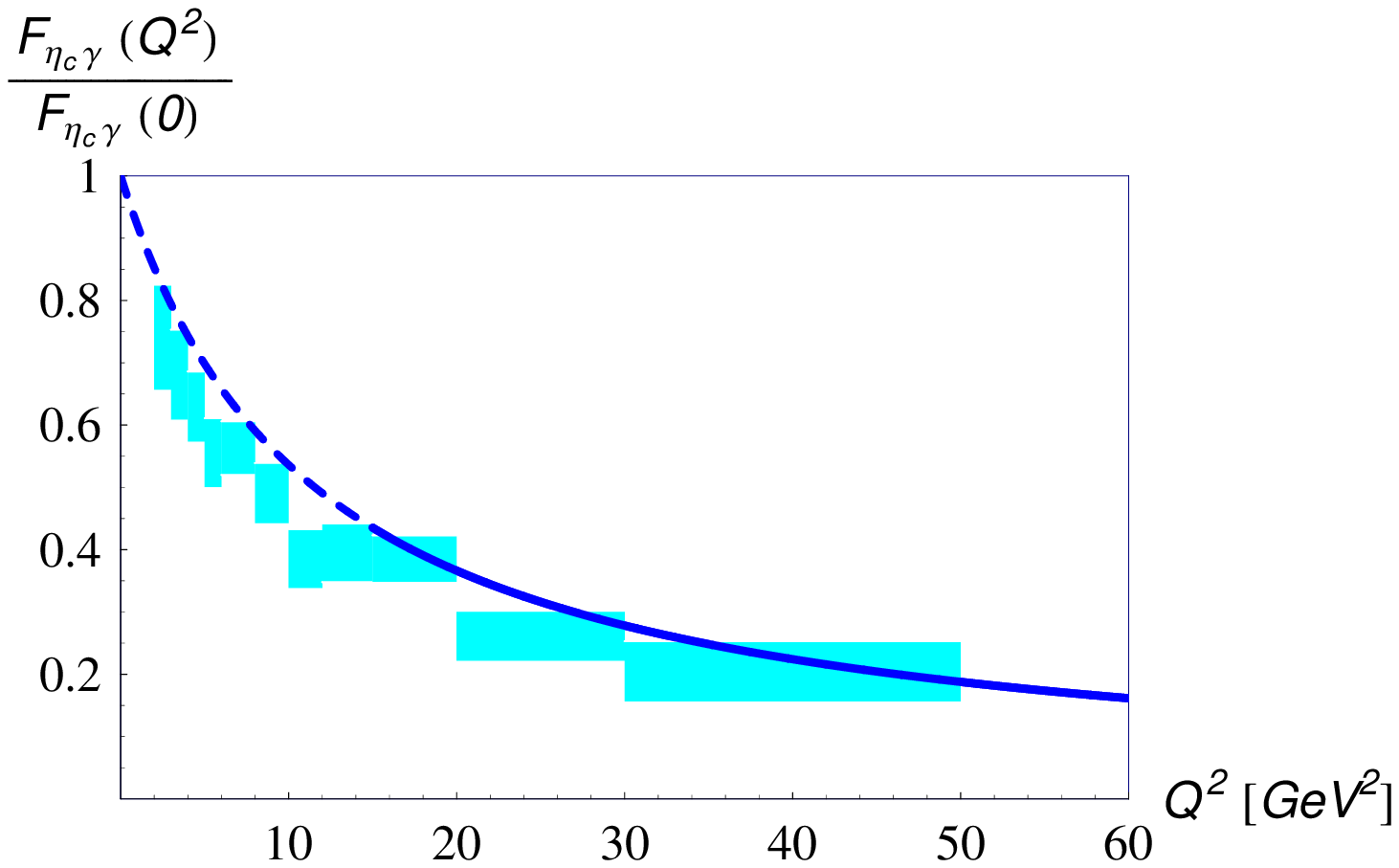}\\(c)&(d)
\end{tabular}\caption{Form factor for the transition
$\gamma\,\gamma^*\to\eta_c$: exact effective thresholds $s_{\rm
eff}^{AVV}(Q^2\to\infty,\beta)$ (a) and $s_{\rm
eff}^{PVV}(Q^2\to\infty,\beta)$ (b); form factors obtained for
finite $Q^2$ from the LD sum rules for the correlators $\langle
AVV\rangle$ and $\langle PVV\rangle$ (c); LD model for the
correlator $\langle PVV\rangle$ confronted with experimental data
by {\sc BaBar}~\cite{babar2010} (d).}\label{Fig:2}\end{figure}

In the case of massive quarks, we may exploit not only the
correlation function $\langle AVV\rangle$ as in Eq.~(\ref{AVV})
but also the correlation function $\langle PVV\rangle$
\cite{lm2011}. For each of these objects, an LD model may be
constructed. By matching to the pQCD factorization formula, we
derive $s_{\rm eff}(Q^2\to\infty,\beta)$ for $\langle AVV\rangle$
and for $\langle PVV\rangle$. The results of the corresponding
calculation for $\eta_c$ are depicted in Fig.~\ref{Fig:2}.
Obviously, the exact effective thresholds corresponding to
$\langle AVV\rangle$ and $\langle PVV\rangle$, $s_{\rm
eff}^{AVV}(Q^2\to\infty,\beta)$ and $s_{\rm
eff}^{PVV}(Q^2\to\infty,\beta),$ differ from each other; they also
differ from the effective thresholds of the relevant two-point
correlation functions.

\emph{Assuming\/} that $s_{\rm eff}(Q^2,\beta)=s_{\rm
eff}(Q^2\to\infty,\beta)$, we obtain the results shown in
Fig.~\ref{Fig:2}. For the above reasons, at very small $Q^2$ the
applicability of our LD model is not guaranteed. Nevertheless,
applying our LD model down to $Q^2=0$ predicts
$F_{\eta_c\gamma}(0)=0.067\;\mbox{GeV}^{-1}$ from the analysis of
$\langle AVV\rangle$ and
$F_{\eta_c\gamma}(0)=0.086\;\mbox{GeV}^{-1}$ from the analysis of
$\langle PVV\rangle$; this has to be compared with the
experimental number
$F_{\eta_c\gamma}(Q^2=0)=0.08\pm0.01\;\mbox{GeV}^{-1}.$ Seemingly,
the LD model based on the correlator $\langle PVV\rangle$ gives
reliable predictions for a broad range of momentum transfers $Q^2$
starting even at very low values of $Q^2$ (cf.~\cite{kroll}).

\section{$\gamma\,\gamma^*\to(\eta,\eta^{\prime})$ form factors}
\begin{figure}[h!]\begin{tabular}{cc}
\includegraphics[width=7.25cm]{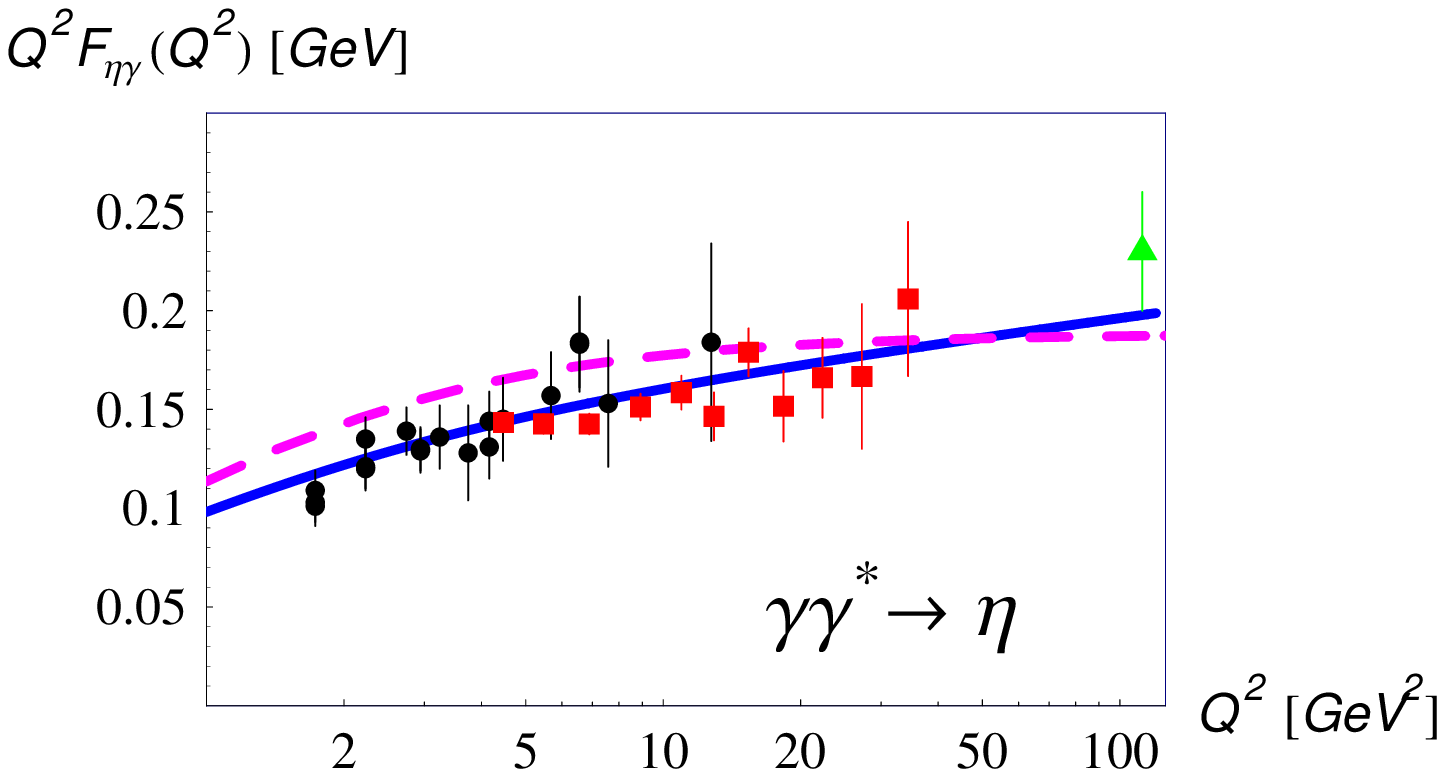}&
\includegraphics[width=7.25cm]{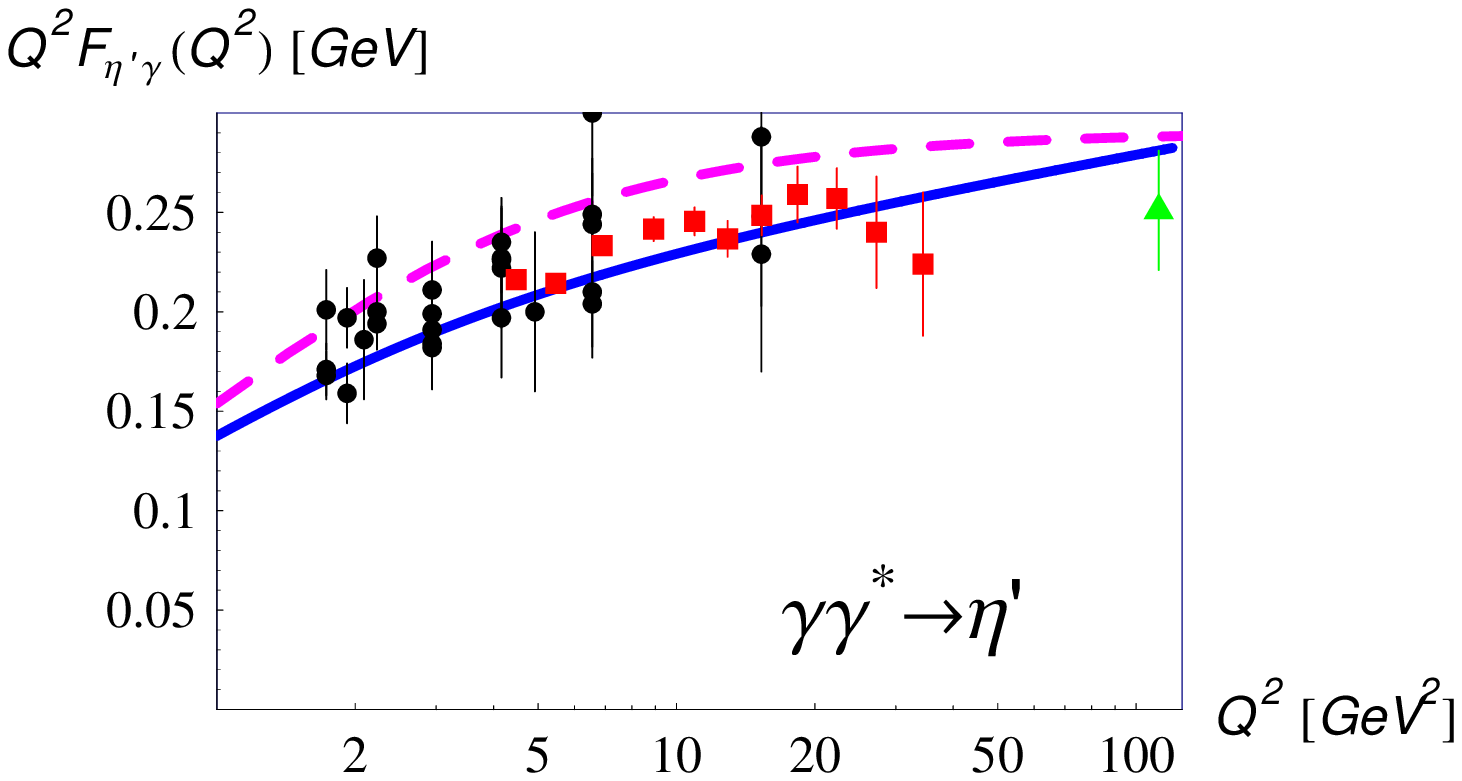}
\end{tabular}\caption{$\gamma\,\gamma^*\to(\eta,\eta^{\prime})$
transition form factors $F_{(\eta,\eta^\prime)\gamma}(Q^2)$: LD
predictions \cite{blm2011,lm2011} (dashed lines) and recent fits
\cite{ms2012} (solid lines) to the experimental data
\cite{cello-cleo,babar1}.}\label{Fig:3}\end{figure}

Here, the mixing of strange and nonstrange components
\cite{mixing} must be taken into account:
\begin{eqnarray}
F_{\eta\gamma}=F_{n\gamma}\cos\phi-F_{s\gamma}\sin\phi,\qquad
F_{\eta'\gamma}=F_{n\gamma}\sin\phi+F_{s\gamma}\cos\phi,\qquad
\phi\approx38^0,
\end{eqnarray}
with $n\to(\bar uu +\bar dd)/\sqrt{2}$ and $s\to\bar ss$. The LD
expressions for these two form factors read
\begin{eqnarray}
F_{n\gamma}(Q^2)=\frac{1}{f_n}\int\limits_0^{s_{\rm
eff}^{(n)}(Q^2)}{\rm d}s\,\Delta_n(s,Q^2),\qquad
F_{s\gamma}(Q^2)=\frac{1}{f_s}\int\limits_0^{s_{\rm
eff}^{(s)}(Q^2)}{\rm d}s\,\Delta_s(s,Q^2).
\end{eqnarray}
Accordingly, two separate effective thresholds emerge: $s_{\rm
eff}^{(n)}=4\pi^2f_n^2$, $s_{\rm eff}^{(s)}=4\pi^2f_s^2$, with
$f_n\approx1.07f_\pi$, $f_s\approx1.36f_\pi$. The outcomes from
the LD model \cite{blm2011,lm2011} and the experimental data
\cite{cello-cleo,babar1} are in reasonable agreement with each
other (Fig.~\ref{Fig:3}).

\section{$\gamma\,\gamma^*\to\pi^0$ form factor}

\begin{figure}[hbt]\begin{tabular}{cc}
\includegraphics[width=7.23cm]{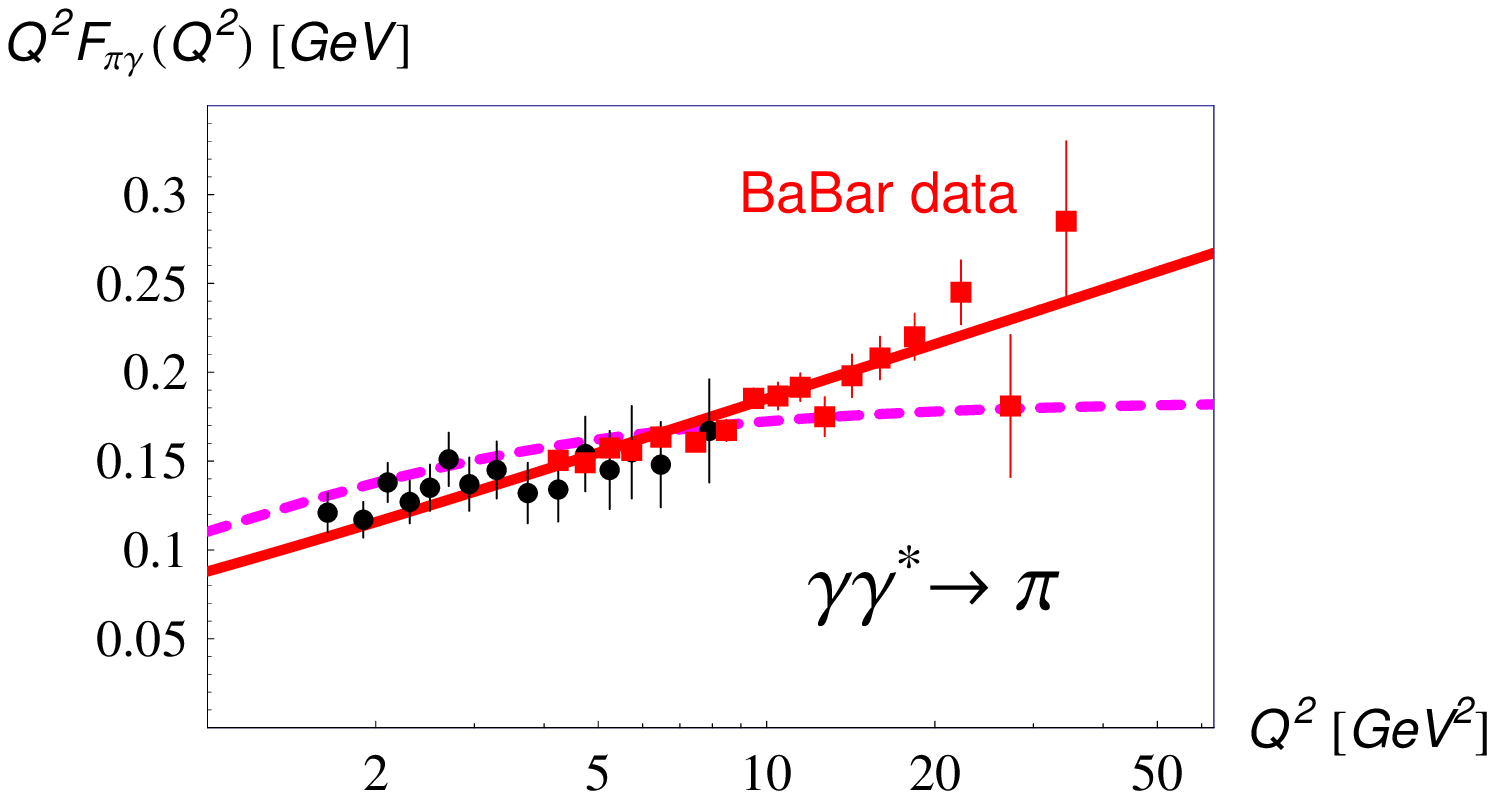}&
\includegraphics[width=7.23cm]{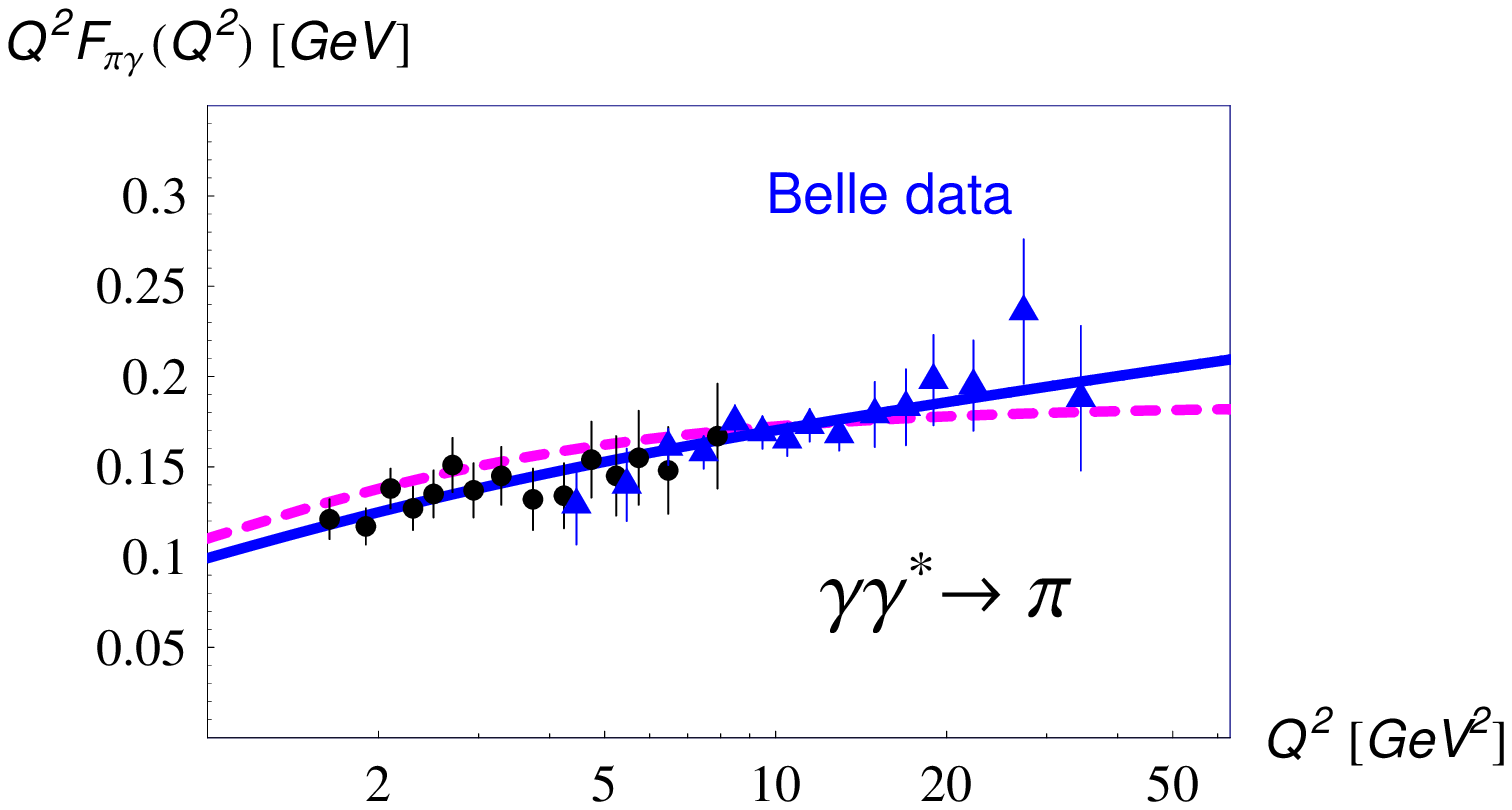}\\
\includegraphics[width=7.23cm]{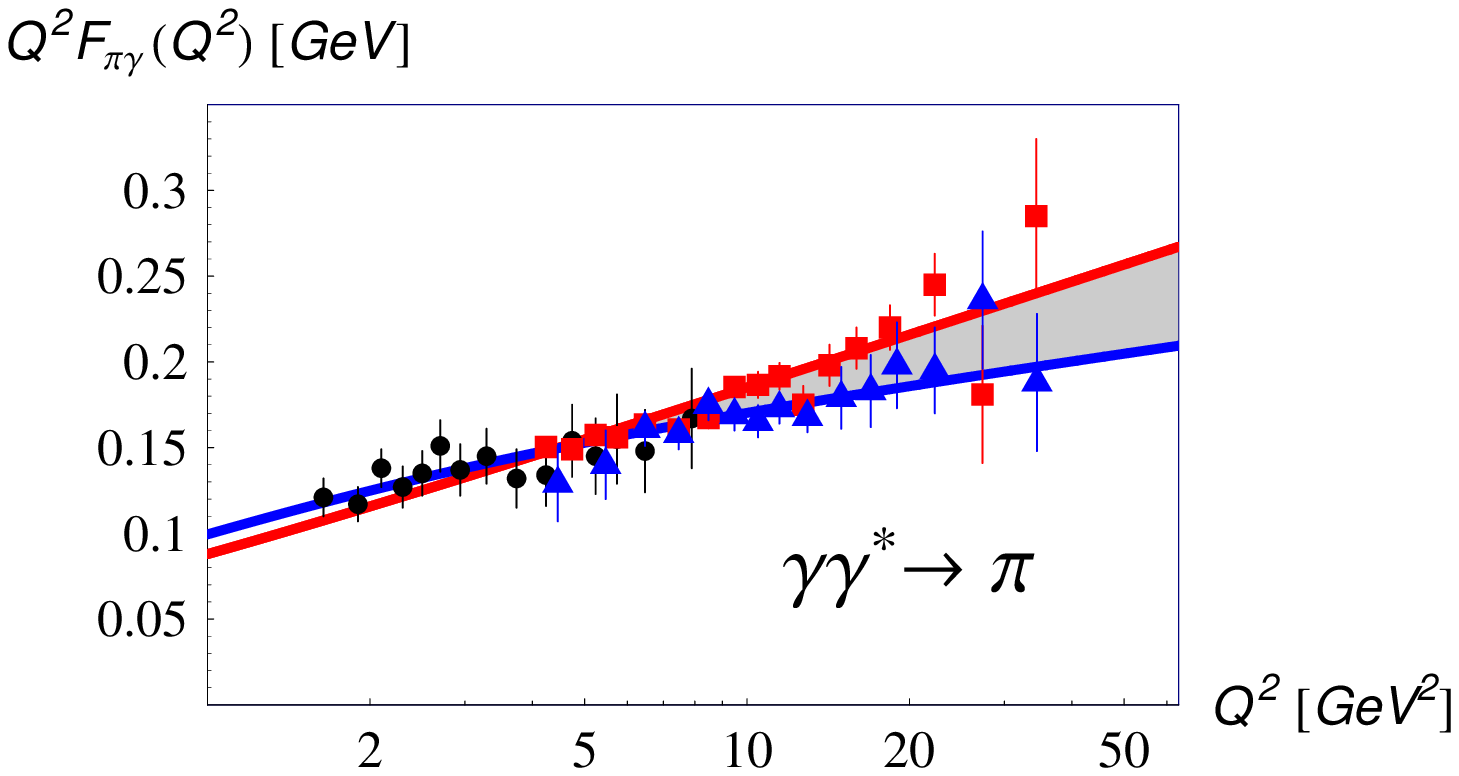}&
\includegraphics[width=7.23cm]{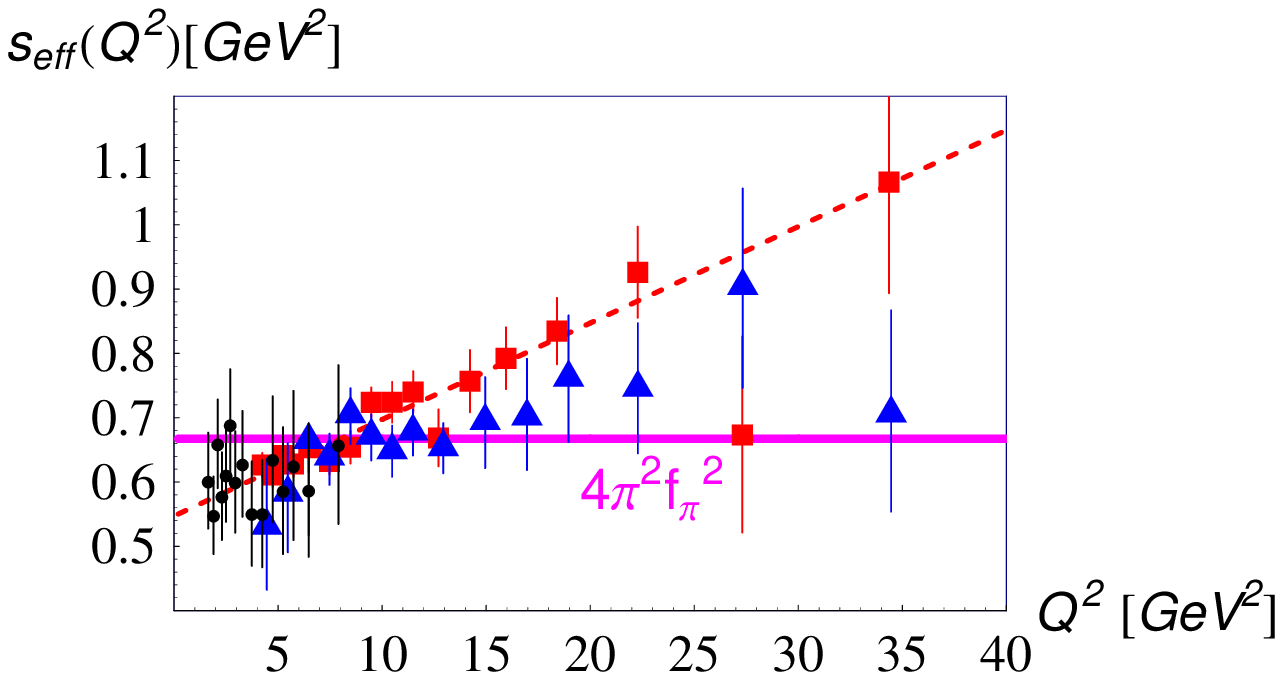}\end{tabular}
\caption{$\gamma\,\gamma^*\to\pi^0$ transition form factor
$F_{\pi\gamma}(Q^2)$: LD prediction (magenta lines) and a fit
\cite{ms2012} (solid lines) to the data
\cite{cello-cleo,babar,belle}. The equivalent effective threshold
$s_{\rm eff}(Q^2)$ for each data point is found via
(\ref{srfp}).}\label{Fig:4}\end{figure}

First of all, we emphasize that the large-$Q^2$ behaviour of the
$\eta$, $\eta'$, and $\pi^0$ form factors is determined by the
spectral densities of perturbative QCD diagrams and should
therefore be the same for all light pseudoscalars \cite{ms2012}.
In order to demonstrate this, we observe that the sum rule for
$\langle AVV\rangle$ in the LD limit $\tau=0$ is equivalent to the
anomaly sum rule \cite{teryaev2}
\begin{equation}\label{pigamma}
F_{\pi\gamma}(Q^2)=\frac{1}{2\sqrt{2}\,\pi^2f_\pi}
\left[1-2\pi\int\limits_{s_{\rm th}}^\infty{\rm
d}s\,\Delta^{I=1}_{\rm cont}(s,Q^2)\right].
\end{equation}
Similar relations arise for the $I=0$ and the $\bar ss$ channels.
As shown in Ref.~\cite{ms2012}, the form factors
$F_{\pi\gamma}(Q^2)$, $F_{\eta\gamma}(Q^2)$, and
$F_{\eta'\gamma}(Q^2)$ at large $Q^2$ are determined by the
behaviour of the appropriate $\Delta_{\rm cont}(s,Q^2)$ at large
$s$. By quark--hadron duality, the latter are equal to the
corresponding $\Delta_{\rm pQCD}(s,Q^2)$; these are purely
perturbative quantities and therefore~equal to each other for the
different channels.

However, the {\sc BaBar} data for the $\pi$ transition form factor
exhibit a clear disagreement with both the $\eta$, $\eta'$ form
factors and the LD model at $Q^2$ as large as $40\;\mbox{GeV}^2$.
Moreover, opposite to findings in quantum mechanics, the
violations of LD rise with $Q^2$ even in the region
$Q^2\approx40\;\mbox{GeV}^2$! We thus conclude that the {\sc
BaBar} results are hard to understand in QCD (see also
\cite{findings}). Noteworthy, recent Belle measurements of the
$\pi\gamma$ form factor---although being statistically consistent
with the {\sc BaBar} findings (see \cite{agaev,pere})---are fully
compatible with the $\eta$ and $\eta'$ data as well as with the
onset of the LD regime already in the region
$Q^2\ge5$--$10\;\mbox{GeV}^2$, in full agreement with our
quantum-mechanical experience.

\section{Conclusions}
We studied the $\pi^0$, $\eta$, $\eta'$, and $\eta_c$ transition
form factors by QCD sum rules in LD limit; the key parameter---the
effective continuum threshold---was determined by matching the LD
form factors to QCD factorization formulas. Our main conclusions
are the following:

\noindent $\bullet$ For all $P\to\gamma\,\gamma^*$ form factors
studied, the LD model should work well in a region of $Q^2$ larger
than a few GeV$^2$: the LD model works reasonably well for the
$\eta\to\gamma\,\gamma^*$, $\eta'\to\gamma\,\gamma^*$, and
$\eta_c\to\gamma\,\gamma^*$ form factors. For
$\pi^0\to\gamma\,\gamma^*$, the {\sc BaBar} data indicate an
extreme violation of local duality, prompting a linearly rising
(instead of a constant) effective threshold. In contrast to this,
the Belle data exhibit an agreement with the predictions of the
LD~model.

\noindent $\bullet$ Nevertheless, a better fit to the full set of
the meson--photon form-factor data seems to prefer a small
logarithmic rise of $Q^2F_{P\gamma}(Q^2)$ \cite{ms2012}. If
established experimentally, this rise would require the presence
of a $1/s$ duality-violating term in the ratio of the hadron~and
the QCD spectral densities.

\noindent $\bullet$ A high accuracy of the LD model has
implications for the pion's \emph{elastic\/} form factor: we can
show that the accuracy of the LD model for the \emph{elastic\/}
form factor increases with~$Q^2$ in the region
$Q^2\approx4$--$8\;\mbox{GeV}^2$ \cite{blm2011}. The accurate data
on the pion form factor suggest that the LD limit for the
effective threshold, $s_{\rm eff}(Q^2\to\infty)=4\pi^2f_\pi^2$,
may be reached already at $Q^2=5$--$6\;\mbox{GeV}^2$. This
property should be testable with the JLab upgrade CLAS12.

\begin{theacknowledgments}D.M.\ was supported by the Austrian
Science Fund (FWF) under project no.~P22843.
\end{theacknowledgments}

\bibliographystyle{aipproc}
\begin{thebibliography}{99}
\bibitem{cello-cleo}H.~J.~Behrend \emph{et al.\/},
\emph{Z.~Phys.~C\/} \textbf{49}, 401 (1991); J.~Gronberg \emph{et
al.\/}, \emph{Phys.~Rev.~D\/} \textbf{57}, 33 (1998).
\bibitem{babar}B.~Aubert \emph{et al.\/}, \emph{Phys.~Rev.~D\/}
\textbf{80}, 052002 (2009).
\bibitem{babar2010}J.~P.~Lees \emph{et al.\/},
\emph{Phys.~Rev.~D\/} \textbf{81}, 052010 (2010).
\bibitem{babar1}P.~del Amo Sanchez \emph{et al.\/},
\emph{Phys.~Rev.~D\/} \textbf{84}, 052001 (2011).
\bibitem{belle}S.~Uehara \emph{et al.\/}, arXiv:1205.3249.
\bibitem{brodsky}G.~P.~Lepage and S.~J.~Brodsky,
\emph{Phys.~Rev.~D\/} \textbf{22}, 2157 (1980).
\bibitem{blm2011}V.~Braguta, W.~Lucha, and D.~Melikhov,
\emph{Phys.~Lett.~B\/} \textbf{661}, 354 (2008); I.~Balakireva,
W.~Lucha, and D.~Melikhov, \emph{J.~Phys.~G\/} \textbf{39}, 055007
(2012) [arXiv:1103.3781]; \emph{Phys.~Rev.~D\/} \textbf{85},
036006 (2012); \emph{Phys.~Atom.~Nucl.\/} \textbf{75} (2012) (in
press) [arXiv:1203.2599].
\bibitem{lm2011}W.~Lucha and D.~Melikhov, \emph{J.~Phys.~G\/}
\textbf{39}, 045003 (2012) [arXiv:1110.2080];
\emph{Phys.~Rev.~D\/} \textbf{86}, 016001 (2012)
[arXiv:1205.4587].
\bibitem{1loop}J.~Ho\v rej\v s\' i and O.~V.~Teryaev,
\emph{Z.~Phys.~C\/} \textbf{65}, 691 (1995); D.~Melikhov and
B.~Stech, \emph{Phys.~Rev.~Lett.\/}~{\bf 88}, 151601 (2002);
D.~Melikhov, \emph{Eur.~Phys.~J.~direct\/} {\bf C4}, 2 (2002)
[arXiv:hep-ph/0110087].
\bibitem{2loop}F.~Jegerlehner and O.~V.~Tarasov,
\emph{Phys.~Lett.~B\/} \textbf{639}, 299 (2006); R.~S.~Pasechnik
and O.~V.~Teryaev, \emph{Phys.~Rev.~D\/} \textbf{73}, 034017
(2006).
\bibitem{lms1}W.~Lucha, D.~Melikhov, and S.~Simula,
\emph{Phys.~Rev.~D\/} \textbf{76}, 036002 (2007);
\emph{Phys.~Lett.~B\/} \textbf{657}, 148 (2007);
\emph{Phys.~Atom.~Nucl.\/} \textbf{71}, 1461 (2008);
\emph{Phys.~Lett.~B\/} \textbf{671}, 445 (2009); D.~Melikhov,
\emph{Phys.~Lett.~B\/} \textbf{671}, 450 (2009).
\bibitem{lms2}W.~Lucha, D.~Melikhov, and S.~Simula,
\emph{Phys.~Rev.~D\/} \textbf{79}, 096011 (2009);
\emph{J.~Phys.~G\/} \textbf{37}, 035003 (2010) [arXiv:0905.0963];
\emph{Phys.~Lett.~B\/} \textbf{687}, 48 (2010);
\emph{Phys.~Atom.~Nucl.\/} \textbf{73}, 1770 (2010);
\emph{J.~Phys.~G\/} \textbf{38}, 105002 (2011) [arXiv:1008.2698];
\emph{Phys.~Lett.~B\/} \textbf{701}, 82 (2011); W.~Lucha,
D.~Melikhov, H.~Sazdjian, and S.~Simula, \emph{Phys.~Rev.~D\/}
\textbf{80}, 114028 (2009).
\bibitem{ld}V.~A.~Nesterenko and A.~V.~Radyushkin,
\emph{Phys.~Lett.~B\/} \textbf{115}, 410 (1982).
\bibitem{Lucha98}W.~Lucha and F.~F.~Sch\"oberl,
\emph{Int.~J.~Mod.~Phys.~C\/} \textbf{10}, 607 (1999).
\bibitem{kroll}P.~Kroll, \emph{Eur.~Phys.~J.~C\/} {\bf 71}, 1623
(2011).
\bibitem{mixing}V.~V.~Anisovich, D.~I.~Melikhov, and V.~A.~Nikonov,
\emph{Phys.~Rev.~D\/} {\bf 55}, 2918 (1997); V.~V.~Anisovich,
D.~V.~Bugg, D.~I.~Melikhov, and V.~A.~Nikonov,
\emph{Phys.~Lett.~B\/} {\bf 404}, 166 (1997); T.~Feldmann,
P.~Kroll, and B.~Stech, \emph{Phys.~Rev.~D\/} {\bf 58}, 114006
(1998); \emph{Phys.~Lett.~B\/} {\bf 449}, 339 (1999).
\bibitem{ms2012}D.~Melikhov and B.~Stech, \emph{Phys.~Rev.~D\/}
\textbf{85}, 051901 (2012); arXiv:1206.5764.
\bibitem{teryaev2}Y.~N.~Klopot, A.~G.~Oganesian, and O.~V.~Teryaev,
\emph{Phys.~Lett.~B\/} {\bf 695}, 130 (2011);
\emph{Phys.~Rev.~D\/} {\bf 84}, 051901 (2011).
\bibitem{findings}H.~L.~L.~Roberts \emph{et al.\/},
\emph{Phys.~Rev.~C\/} \textbf{82}, 065202 (2010); S.~J.~Brodsky,
F.-G.~Cao, and G.~F.~de T\'eramond, \emph{Phys.~Rev.~D\/}
\textbf{84}, 033001 (2011); \textbf{84}, 075012 (2011);
A.~P.~Bakulev, S.~V.~Mikhailov,~A.~V.\ Pimikov, and
N.~G.~Stefanis, \emph{Phys.~Rev.~D\/} \textbf{84}, 034014 (2011);
arXiv:1205.3770.
\bibitem{agaev}S.~S.~Agaev, V.~M.~Braun, N.~Offen, and
F.~A.~Porkert, \emph{Phys.~Rev.~D\/} {\bf 83}, 054020 (2011);
arXiv: 1206.3968.
\bibitem{pere}P.~Masjuan, arXiv:1206.2549.
\end{thebibliography}
\end{document}